# Chip-integrated van der Waals PN heterojunction photodetector with low dark current and high responsivity

Ruijuan Tian[1], Xuetao Gan[1,*], Chen Li[1], Xiaoqing Chen[1], Siqi Hu[1], Linpeng Gu[1], Dries Van Thourhout[2], Andres Castellanos-Gomez[3], Zhipei Sun[4], Jianlin Zhao[1]

1. Key Laboratory of Light Field Manipulation and Information Acquisition, Ministry of Industry and Information Technology, and Shaanxi Key Laboratory of Optical Information Technology, School of Physical Science and Technology, Northwestern Polytechnical University, 710129 Xi'an, China
2. Photonics Research Group, Center for Nano and Biophotonics, Ghent University, B-9000 Gent, Belgium
3. Materials Science Factory, Instituto de Ciencia de Materiales de Madrid (ICMM-CSIC), E-28049 Madrid, Spain
4. Department of Micro- and Nanosciences, Aalto University, Tietotie 3, FI-00076 Espoo, Finland
*Email: xuetaogan@nwpu.edu.cn

**Abstract**: Two-dimensional materials are attractive for constructing high-performance photonic chip-integrated photodetectors because of their remarkable electronic and optical properties and dangling-bond-free surfaces. However, the reported chip-integrated two-dimensional material photodetectors were mainly implemented with the configuration of metal-semiconductor-metal, suffering from high dark currents and low responsivities at high operation speed. Here, we report a van der Waals PN heterojunction photodetector, composed of p-type black phosphorous and n-type molybdenum telluride, integrated on a silicon nitride waveguide. The built-in electric field of the PN heterojunction significantly suppresses the dark current and improves the responsivity. Under a bias of 1 V pointing from n-type molybdenum telluride to p-type black phosphorous, the dark current is lower than 7 nA, which is more than two orders of magnitude lower than those reported in other waveguide-integrated black phosphorus photodetectors. An intrinsic responsivity up to 577 mA W$^{-1}$ is obtained. Remarkably, the van der Waals PN heterojunction is tunable by the electrostatic doping to further engineer its rectification and improve the photodetection, enabling an increased responsivity of 709 mA W$^{-1}$. Besides, the heterojunction photodetector exhibits a response bandwidth of ~1.0 GHz and a uniform photodetection over a wide spectral range, as experimentally measured from 1500 to 1630 nm. The demonstrated chip-integrated van der Waals PN heterojunction photodetector with low dark current, high responsivity and fast response has great potentials to develop high-performance on-chip photodetectors for various photonic integrated circuits based on silicon, lithium niobate, polymer, etc.

## Introduction

On-chip photonic integrated circuits (PICs) could provide a versatile platform to revolutionize optical computing, communications, chemical-/bio-sensing, LiDARs, etc[1-4]. Currently, a variety of materials are being employed to develop PICs with their own superiorities. For example, silicon PIC is compatible with the mature CMOS technology for low-cost and large-scale production[5]; Silicon nitride PIC could tolerate moderately high optical power and large fabrication errors[6,7]; Lithium niobate PIC could achieve perfect electro-optic modulations with low driven voltage and high linearity[8]. One of the handicaps in these PICs, however, is the monolithic integration of waveguides and photodetectors with a single material. To support the light transmission in the waveguide, the PIC materials cannot absorb the optical signal, making it impossible to realize the integrated photodetector out of a single material. To solve this, hetero-integrations of absorptive bulk materials on PICs have been implemented, which though still present open challenges. For silicon PICs, the heteroepitaxy of germanium or III-V compound semiconductors has achieved decent integrated photodetectors[9-11]. Despite continued advancement, technological factors limit the widespread usage of such detectors, specifically including the high costs associated with the processing of germanium or III-V semiconductors. Wafer-bonding technique is an alternative, which unfortunately involves complicated processes and has material interface issues. For PICs of dielectric materials, such as silicon nitride and lithium niobate, bulk semiconductors can hardly be epitaxially grown on top, though only few works about the integrated photodetectors were reported with complicated bonding processes[12-14].

Recently, two-dimensional (2D) materials have emerged as an attractive photon-absorption material for chip-integrated photodetectors[15-20]. 2D materials have no surface dangling bonds, which eliminates the lattice-mismatch constraints to hetero-integrate them with PICs. The family of 2D materials has a rich variety of electronic and optical properties, including semi-metallic graphene, insulating boron nitride, semiconducting transition metal dichalcogenides and black phosphorus. As a consequence, chip-integrated photodetectors operating at various spectral ranges could be constructed by choosing appropriate 2D materials. Optoelectronic properties of 2D materials could be engineered further by stacking dissimilar 2D materials to create van der Waals heterostructures. In addition, large-scale growth of 2D materials via chemical vapor deposition method has also been demonstrated with the simplicity of device fabrications[21]. In the last few years, chip-integrated 2D material photodetectors have been reported with excellent merits. Graphene photodetectors were integrated on waveguides of silicon, silicon nitride, and chalcogenide glass[21-24]. An operation bandwidth exceeding 110 GHz was realized owing to the ultrahigh carrier mobility of the graphene[25]. A monolayer molybdenum disulfide was integrated on a silicon nitride waveguide for the photodetector on visible PIC[26], which exhibits a remarkable responsivity of $10^3$ A W$^{-1}$. However, these chip-integrated 2D material photodetectors still have gaps for future PICs, which require a photodetection performance with the combination of low dark current, high responsivity, and fast response speed. Graphene's zero bandgap causes high dark currents in its photodetectors. Photodetectors of other 2D materials were normally carried out with the geometry of metal-semiconductor-metal (M-S-M), which results in high dark current and low operation bandwidth (see Supplementary Information for more details).

Here, we report that integrating van der Waals PN heterojunctions of 2D materials on optical waveguides can provide a promising strategy to realize chip-integrated photodetectors with low dark current, high responsivity, and fast speed. In those photodetectors based on bulk materials, PN junctions were widely employed to achieve high performance. Their rectifying characteristics allow the suppression of dark current and the improvements of responsivity and operation bandwidth. PN junctions of bulk materials are generally realized by complicated ion implantation, diffusion of dopants, or epitaxy. Differently, the PN heterojunction of 2D materials employed in this work is formed by simply van der Waals stacking few-layer black phosphorus (BP) and molybdenum ditelluride (MoTe$_2$), which are naturally p- and n-doped, respectively. Because of the ultrathin thickness and a steep interfacial charge carrier gradient, this van der Waals PN heterojunction allows the electrostatic doping to engineer its rectification behavior and to improve the photodetection performance. The BP/MoTe$_2$ PN heterojunction is integrated on a silicon nitride waveguide to carry out the chip-integrated photodetector. Under an external bias voltage of 1 V pointing from n-type MoTe$_2$ to p-type BP, the dark current is lower than 7 nA, which is more than two orders of magnitude lower than those reported in other waveguide-integrated BP photodetectors. The intrinsic responsivity is about 600 mA W$^{-1}$. The dynamic response presents a 3 dB-bandwidth of ~1.0 GHz. Moreover, the operation over a broad spectral range is obtained due to BP's intrinsic narrow bandgap. The proposed van der Waals PN heterojunction could be constructed by other 2D materials[27,28] for photodetectors in other desirable spectral ranges on PICs.

## Results

### Structure of the waveguide-integrated van der Waals PN heterojunction

Figure 1a presents the schematic of the waveguide-integrated BP/MoTe$_2$ heterojunction photodetector. As reported in our previous work, by mechanically stacking p-doped BP and n-doped MoTe$_2$, their interface forms a van der Waals PN heterojunction with a vertical built-in electric field[29]. The integration of this heterostructure on a silicon nitride waveguide enables its coupling with the evanescent field of the guiding mode. The light absorption is enhanced effectively due to the extended light-BP/MoTe$_2$ interaction length. Drain and source electrodes are transferred directly on the BP and MoTe$_2$ along the two sides of the waveguide, respectively[30] (see Materials and methods section). Note, though the direct transfer-printing technique is employed to fabricate the metal electrodes, the ordinary method with metal deposition and lift-off is also valid for the proposed integrated photodetector[16,20]. The carrier transport direction in the BP/MoTe$_2$ heterojunction channel is perpendicular to the light propagation direction in the waveguide. For light transmitting in the waveguide at the telecom wavelength range, the BP layer would absorb it to generate photocarriers. With the built-in electric field in the PN heterojunction or the external electric field from an applied bias, the photocarriers would be rapidly separated at the heterostructure interface, which yields photocurrent across the heterostructure channel. The MoTe$_2$ layer can also contribute to the photocurrent if the light has a shorter wavelength, such as the visible light, making the waveguide-integrated photodetector more efficient. Actually, other n-doped 2D materials, such as MoS$_2$ and InSe, could be alternative for replacing the n-doped MoTe$_2$ layer if the clean interface of the van der Waals heterostructure is guaranteed. Here, a layer of hexagonal boron nitride (h-BN) is placed between the BP/MoTe$_2$ heterostructure and the silicon nitride waveguide to minimize the substrate rough surface and local potential puddles, which might degrade the intrinsic electronic properties of the 2D materials. Since BP is unstable in the ambient condition, another h-BN layer is employed to encapsulate the BP/MoTe$_2$ heterostructure, which could ensure the stability of device performance during the testing.

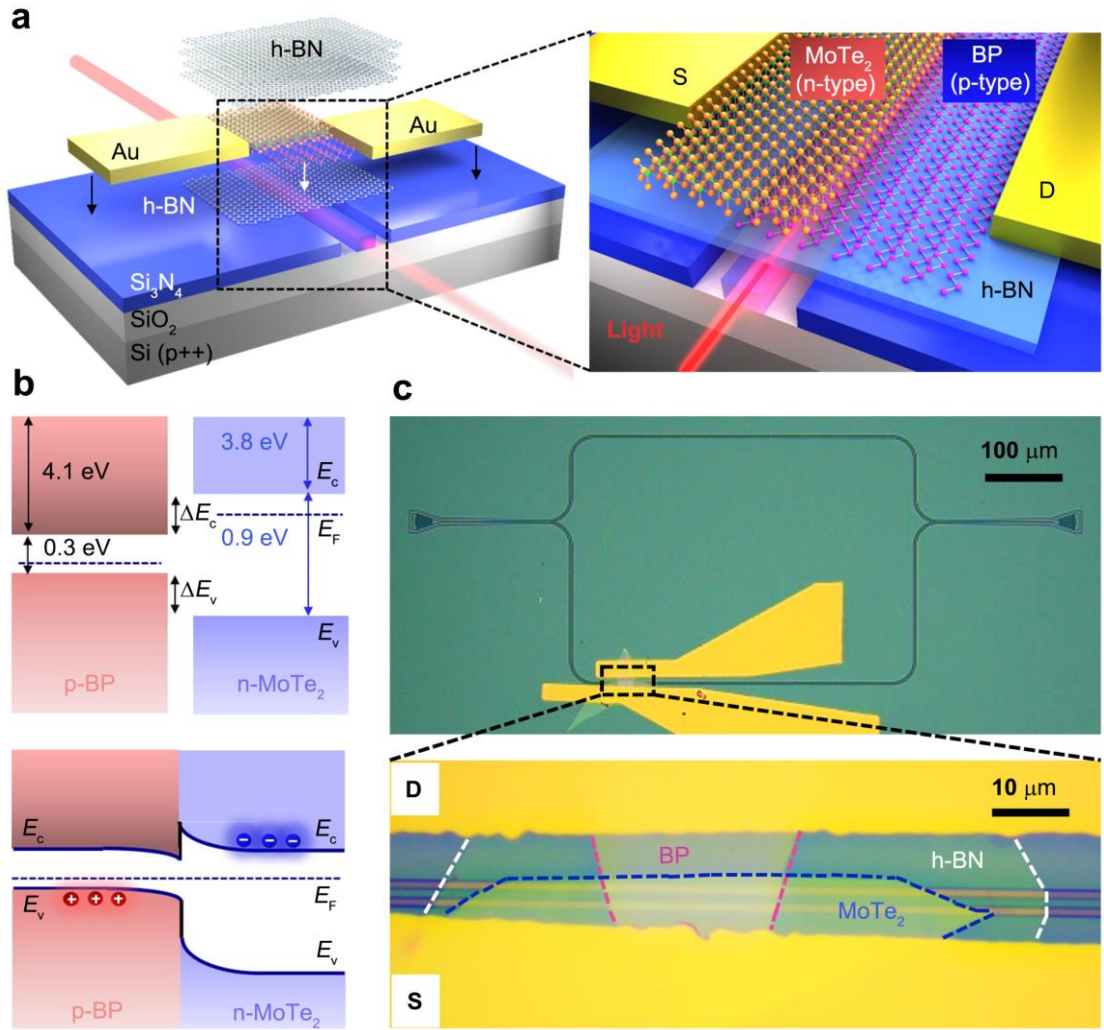

**Fig. 1 BP/MoTe$_2$ van der Waals PN heterojunction photodetector integrated on a silicon nitride waveguide**. **a** Schematic of the waveguide-integrated van der Waals PN heterojunction photodetector composed of stacking p-doped BP and n-doped MoTe$_2$, which are evanescently coupled with the guiding mode of the silicon nitride waveguide. **b** Top: Band profiles of BP and MoTe$_2$ layers in the non-equilibrium state. Bottom: Band alignment of BP/MoTe$_2$ PN heterojunction in the thermal equilibrium state. **c** Top: Optical microscope image of the fabricated device, where a BP/MoTe$_2$ heterojunction is integrated on one arm of a Mach-Zehnder interferometer. Bottom: Zoom-in view of the device indicated by the black square box in the top panel.

The employed silicon nitride waveguide is fabricated in a silicon nitride slab (with a thickness of 300 nm) on a silicon substrate with a middle silicon oxide layer (with a thickness of 2 μm). The silicon substrate is highly doped, which could be functioned as a global back gate of the BP/MoTe$_2$ heterostructure. Similar to the configuration of a field effect transistor, the metal electrodes contacting the BP and MoTe$_2$ layers could be defined as the drain and source electrodes, respectively. With the buried silicon oxide layer and the top silicon nitride layer as the dielectric layer, a vertical electric field could be actuated over the top BP/MoTe$_2$ heterostructure when a gate voltage is applied between the drain electrode and the bottom doped silicon substrate. The carrier densities in the BP/MoTe$_2$ heterostructure could be tuned by the electrostatic field, resulting in the tunable heterojunctions.

Figure 1b shows the energy band diagrams of the BP/MoTe$_2$ van der Waals heterostructure in non-equilibrium (the top panel) and thermal equilibrium states (the bottom panel). The bandgap and electron affinity of BP (MoTe$_2$) are ~0.3 eV (~0.9 eV) and ~4.1 eV (~3.8 eV), respectively[31,32]. The differences in bandgaps and electron affinities give rise to the BP/MoTe$_2$ heterojunction with a conduction band offset $\Delta E_c$ ~0.3 eV and a valence band offset $\Delta E_v$ ~0.3 eV. Under the thermal equilibrium state, the interface between p-doped BP and n-doped MoTe$_2$ forms a typical PN diode, as illustrated in the bottom panel of Fig. 1b. The built-in electric potential of the BP/MoTe$_2$ van der Waals heterostructure is simply estimated by the difference between the Fermi levels of p-doped BP and n-doped MoTe$_2$, which is accompanied by a built-in electric field pointing from MoTe$_2$ to BP.

To determine the intrinsic responsivity and internal quantum efficiency of the photodetector, the absolute light absorption on the waveguiding mode by the BP/MoTe$_2$ heterostructure should be quantitatively extracted. To obtain that, a Mach-Zehnder interferometer (MZI) based on the silicon nitride waveguide is utilized, as shown in Fig. 1c. The BP/MoTe$_2$ heterostructure is integrated on one arm of the MZI. The width of the waveguide is 1080 nm to ensure the low-loss single mode transmission around the telecom wavelength range. To facilitate the light coupling between single mode fibers and the silicon nitride MZI, two grating couplers are designed at the ends of the MZI. The MZI patterns are carved out of the silicon nitride slab by electron beam lithography and plasma dry-etching. The bottom h-BN, BP, MoTe$_2$, and top h-BN layers are dry transferred and mechanically stacked onto the silicon nitride waveguide in sequence[29]. Before the final encapsulation of the top h-BN layer, pre-patterned Au pads deposited on a silicon substrate are peeled and mechanically transferred onto the BP and MoTe$_2$ flakes to form the drain and source electrodes, respectively[29,30]. To release the air bubbles between the 2D materials and metal electrodes for their atomic level contact, the finished device is finally annealed in a tube furnace filled with a forming gas (95% Ar and 5% $H_2$). This effectively assists the formation of the BP/$MoTe_2$ PN heterojunction and decent metal contacts. In the fabricated device, the lengths of the BP and $MoTe_2$ flakes overlapping with the waveguide are ~19.77 μm and ~56.6 μm, respectively. Their heterostructure is confirmed by Raman spectroscopy (see Fig. S1). The thicknesses of the BP and $MoTe_2$ layers are evaluated by atomic force microscope and found to be ~13 nm and 10.6 nm (see Fig. S2).

Optical transmissions of the fabricated MZI are measured by coupling a wavelength-tunable narrowband laser (1500 nm to 1630 nm) into one of the grating couplers, and monitoring the output powers at the other grating coupler. Though both the transverse electric (TE) and transverse magnetic (TM) modes are supported in the waveguide, the waveguiding mode with the TE polarization has much lower transmission loss than the TM mode. In our measurement, the TE mode of the silicon nitride waveguide is experimentally excited by controlling the polarization of the incident light from the input fiber to achieve the maximized output power. The transmission spectra could be obtained by sweeping the laser wavelength. During the integration of the BP/$MoTe_2$ heterostructure, the transmission spectra of the MZI are recorded in each step, as shown in Fig. S3. Interference fringes are observed from the transmission spectra with the period determined by the phase difference between the two arms of the MZI. After each transfer process of the 2D materials, the whole interference fringe pattern red-shifts due to the increased phase delay on the long arm of the MZI. Because each integration of 2D materials would induce absorptions or mode scatterings, extinction ratios (ERs) of the interference fringes reduce gradually. From the extracted ERs, the absorption coefficient of the BP layer is evaluated as ~0.09815 dB $μm^{-1}$. The measured light absorption coefficient agrees well with the theoretical prediction value of 0.0911 dB $μm^{-1}$ (See Supplementary Information for detail). In other words, the integrated 19.77 μm long BP layer absorbs about 42.35% of the optical power transmitted in the waveguide. This high light absorption by the BP layer could be attributed to its narrow direct bandgap and the extended interaction length with the optical guiding mode. It could facilitate the high-efficiency in the photodetection.

**Electrical properties and photoresponses of the waveguide-integrated**

## BP/MoTe$_2$ PN heterojunction

Before we couple light into the waveguide covered with the BP/MoTe$_2$ heterostructure, we measure the electrical characteristics of the BP/MoTe$_2$ channel by applying the drain-source bias voltage $V_{DS}$ from −1 V to 1 V to evaluate the performance of the van der Waals PN heterojunction. The positive (negative) $V_{DS}$ means the drain electrode has a higher (lower) electric potential than the source electrode. As shown in the black curve of Fig. 2a, without optical illumination, the drain-source current $I_{DS}$ presents a strong rectifying characteristic under the positive and negative bias voltage $V_{DS}$. The ON/OFF current ratio of this diode behavior is about $1.6 \times 10^3$, and the ideality factor is estimated as 1.87 using the Shockley diode equation[33,34]

$$I_{DS} = \frac{nV_T}{R_s} W \left[ \frac{I_0 R_s}{nV_T} \exp\left( \frac{V_{DS}}{nV_T} \right) \right]$$

where $V_T = k_B T/q$ is the thermal voltage at temperature $T$, $k_B$ is the Boltzmann constant, $q$ is the electron charge, $I_0$ is the saturable current under negative $V_{DS}$, $n$ is the diode ideality factor, $W$ is the Lambert W function and $R_s$ is the series resistance. The result indicates the van der Waals PN heterojunction is successfully formed at the interface of stacked p-doped BP and n-doped MoTe$_2$.

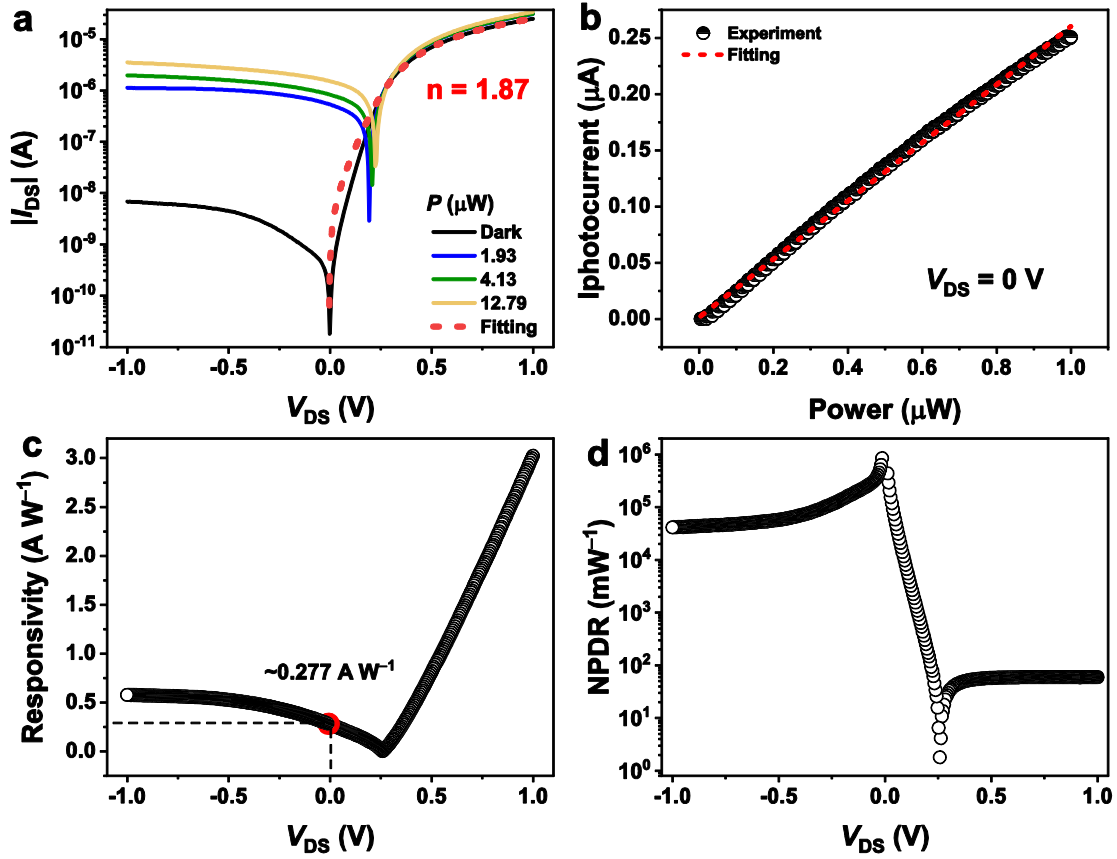

**Fig. 2 Electrical characteristics and photoresponses (with the coupling light at 1503 nm) of the waveguide-integrated BP/MoTe$_2$ heterojunction photodetector. a** Measured current–voltage characteristics of the photodetector without (black line) and with optical illuminations at different BP absorption powers. **b** Measured zero-bias photocurrents versus the absorbed optical powers. **c** Intrinsic responsivity as a function of the bias voltage. **d** Normalized photocurrent-dark-current ratio (NPDR) as a function of the bias voltage.

By coupling light at the wavelength of 1503 nm into the MZI via the grating coupler, the $I_{DS}$–$V_{DS}$ curves of the PN heterojunction are measured at different optical absorption powers by the BP layer, as shown in Fig. 2a. Due to photocarriers in the BP layer, $I_{DS}$ increases considerably under optical illumination, which implies the pronounced photoresponse. The magnitude of the photocurrent $I_{photo}$ is defined by the difference between $I_{DS}$ measured with and without optical illumination. It increases with higher absorbed optical powers because more photocarriers are generated and collected. With the zero bias ($V_{DS}$ = 0 V), owing to the built-in electric field formed at the BP/MoTe$_2$ interface, the heterojunction could separate the photocarriers and generate large photocurrent. It promises the self-powered on-chip photodetection.

Upon illumination, the internal electric field at the BP/MoTe$_2$ heterojunction separates photogenerated carriers and gives rise to a photocurrent at zero external bias (short-circuit current, $I_{sc}$) and a photovoltage with no current flowing (open-circuit voltage, $V_{oc}$). Figure 2b plots the measured $I_{sc}$ as the BP absorption power increases, showing a linear dependence. It confirms the photovoltaic effect of the van der Waals PN heterojunction. In addition, an open-circuit voltage $V_{oc}$ of 228 mV is measured from the heterojunction with a BP absorption power of 12.79 µW, as shown in Fig. S5. The intrinsic responsivity $R_{int}$ is estimated as 277 mA W$^{-1}$, which is defined as $R_{int}$ =$I_{photo}/P_{photo\_int}$. Here, $P_{photo\_int}$ is the optical power absorbed by the BP layer, which is experimentally determined from the interferometry results of the MZI before and after the integration of BP (see the Supplementary Information)[16]. To the best of our knowledge, the obtained zero-bias responsivity is highest among those reported in self-powered waveguide-integrated 2D material photodetectors[15,17,18,21,35]. This could be attributed to the built-in electric field at the interface of the PN heterojunction. The high intrinsic responsivity corresponds to a high internal quantum efficiency $\eta_{IQE}$ of 22.5% at the wavelength of 1503 nm, which is defined as $\eta_{IQE} = R_{int} \times \hbar\omega/q$. Here, $\hbar$ is the reduced Planck constant, $\omega$ is the light angular frequency and $q$ is the elementary charge. The high open-circuit voltage and short-circuit current indicate that a van der Waals PN junction is successfully formed in the waveguide-integrated BP/MoTe$_2$ heterostructure.

We further characterize the photoresponses of the device at varied electrical biases of $V_{DS}$. The extracted photoresponsivities are shown in Fig. 2c. With the negative $V_{DS}$, an external electric field pointing from MoTe$_2$ to BP is built at the interface of the PN junction. It would strengthen the built-in electric field of the junction, which therefore improves the separation efficiency of the photocarriers and subsequently increases the responsivity. At $V_{DS} = -1$ V, the intrinsic (extrinsic) responsivity is increased to 577 (283) mA W$^{-1}$. The extrinsic responsivity ($R_{ext}$) is defined as $R_{ext} = I_{photo}/P_{photo\_ext}$, where $P_{photo\_ext}$ is the optical power coupled into the waveguide via the grating coupler. With the positive $V_{DS}$, the responsivity first decreases to the minimum due to the gradual elimination of the built-in electric field, and then increases rapidly with the effective electric field provided by the external bias considering the thinned depletion region. At $V_{DS} = 1$ V, the intrinsic (extrinsic) responsivity is about 3 (1.48) A W$^{-1}$. This could be attributed to the photoconductive gain. As indicated by the side schematic view of the fabricated device (see Fig. S6), the photogenerated electrons in BP must transport a long channel to get collected by the source electrode, which can lead to a large photoconductive gain. Note that the dark current at $V_{DS}$ = 1 V is about 25 μA, as shown in Fig. 2a. Though the responsivity is high at this positive $V_{DS}$, the large dark current would reduce the signal-to-noise ratio and increase power consumption, which degrades its operation potential.

On the contrary, at $V_{DS}$ = −1 V, the dark current of the device $I_{dark}$ is as low as 6.8 nA, promising a high-performance photodetection by considering its high intrinsic responsivity of 577 mA W$^{-1}$. The dark current of 6.8 nA at $V_{DS}$ = −1 V is around three orders of magnitude lower than that of waveguide-integrated graphene photodetectors[17] and two orders of magnitude lower than that of waveguide-integrated BP photodetectors[16,35] (see Supplementary Table 1), which benefits from the PN heterojunction in the proposed van der Waals heterostructure. Normalized photocurrent-dark-current ratio (NPDR) is another important performance indicator of an on-chip photodetector[36], which is expressed as NPDR = $(I_{photo}/I_{dark})/P$, where $P$ is the laser power coupled in the waveguide. Taking the external responsivity into account, the evaluated NPDRs of the device at different bias voltages are plotted in Fig. 2d. At $V_{DS}$ = −1 V, the NPDR exceeds 4.13×10$^4$ mW$^{-1}$, which is larger than those of waveguide-integrated 2D material photodetectors by two to four orders of magnitude[20,37-39]. The NPDR enhancement could be accredited to the effective suppression of the dark current in the PN heterojunction without the sacrifice of the photocurrent.

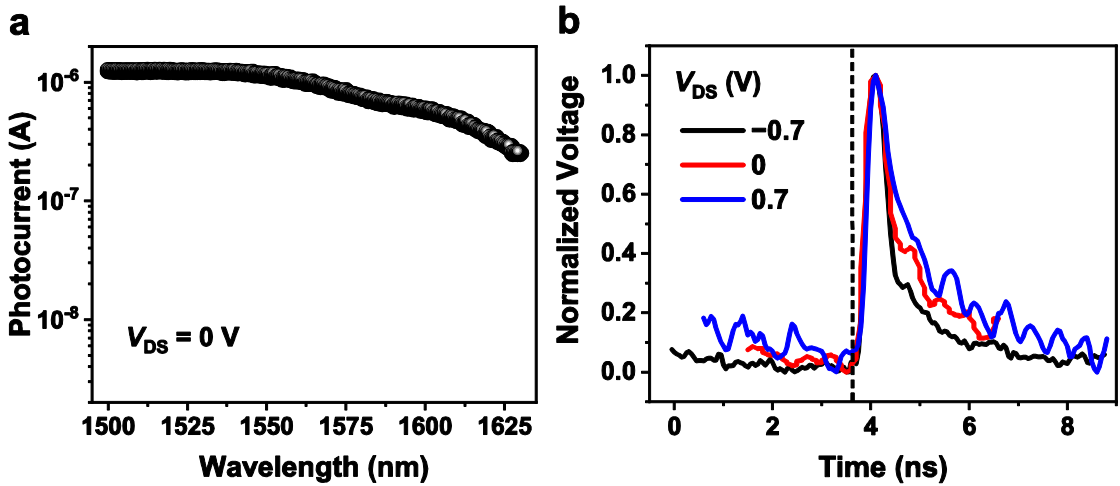

**Fig. 3 Steady and dynamic photoresponses of the waveguide-integrated BP/MoTe₂ PN heterojunction photodetector. a** Photocurrents measured under the optical illumination at the wavelengths ranging from 1500 nm to 1630 nm. **b** Impulse response of the device at zero, positive and negative biases. The vertical dashed line indicates the time point at which the picosecond laser pulse launches.

Because of the intrinsic narrow bandgap of the few-layer BP flake, the constructed BP/MoTe₂ heterojunction on the waveguide could respond to the optical illumination in a wide spectral range, enabling the broadband photodetection. Figure 3a displays the wavelength-dependent photoresponse of the fabricated device at zero bias. By fixing the laser power and sweeping the laser wavelength, photocurrents are obtained over the wavelength range from 1500 to 1630 nm, which is the operation wavelength range of the employed tunable laser. The photocurrents over the measured wavelength range are moderately uniform. As the laser wavelength increases, the gradually decreased photocurrents could be attributed to the degraded coupling efficiency of the grating couplers at the longer wavelength (see Fig. S3c).

To evaluate the dynamic response of this waveguide-integrated van der Waals PN heterojunction photodetector, we carry out measurements of its impulse response. A picosecond pulsed laser at the wavelength of 1550 nm is employed as the incident light source, which provides a train of 5 ps long (full width at half-maximum) optical pulses. By sending the pulses into the device, the generated photocurrent pulses are monitored using a real-time oscilloscope (see Fig. S7). Figure 3b shows the measured impulse responses of the device at zero, positive and negative biases, from which we extract the pulse relaxation time $\Delta t$ by counting the full time width at half-maximum of the pulse peak. The results are $\Delta t = 440$ ps at $V_{DS} = -0.7$ V, $\Delta t = 523.8$ ps at $V_{DS} = 0$ V, $\Delta t = 1157$ ps at $V_{DS} = 0.7$ V. The dynamic response bandwidth can be derived by the time-bandwidth product, giving $f_{3\,dB} = 0.44/\Delta t = $ ~1.0 GHz at $V_{DS} = -0.7$ V, ~0.84 GHz at $V_{DS} = 0$ V, ~0.38 GHz at $V_{DS} = 0.7$ V. Because of the ultrathin junction interface of the vertical van der Waals heterostructure, the carrier diffusion time across it would not limit the response speed. The dependence on the bias voltage of the bandwidth could imply that the operation speed is mainly limited by the resistance and capacitance in the BP/MoTe₂ channel, *i.e.*, the RC constant. With a positive (negative) bias $V_{DS}$, the depleted charge region of the PN BP/MoTe₂ junction becomes narrower (wider), which increases (decreases) the junction capacitance. As a result, the bandwidths of 1.0 GHz and 0.38 GHz are obtained with $V_{DS}$ = −0.7 V and 0.7 V, respectively. Since the absence of instruments to measure the resistance and capacitance of the device at high frequency in our experiment condition, it is difficult to quantitatively analyze their individual influences on the response speed. In the future device, by shortening the channel length and reducing device area of the the BP/MoTe$_2$ heterostructure to only cover on the waveguide, the RC constant could be decreased greatly to further increase operation bandwidth.

## Improving photodetection of the waveguide-integrated BP/MoTe$_2$ PN heterojunction by electrostatic gating

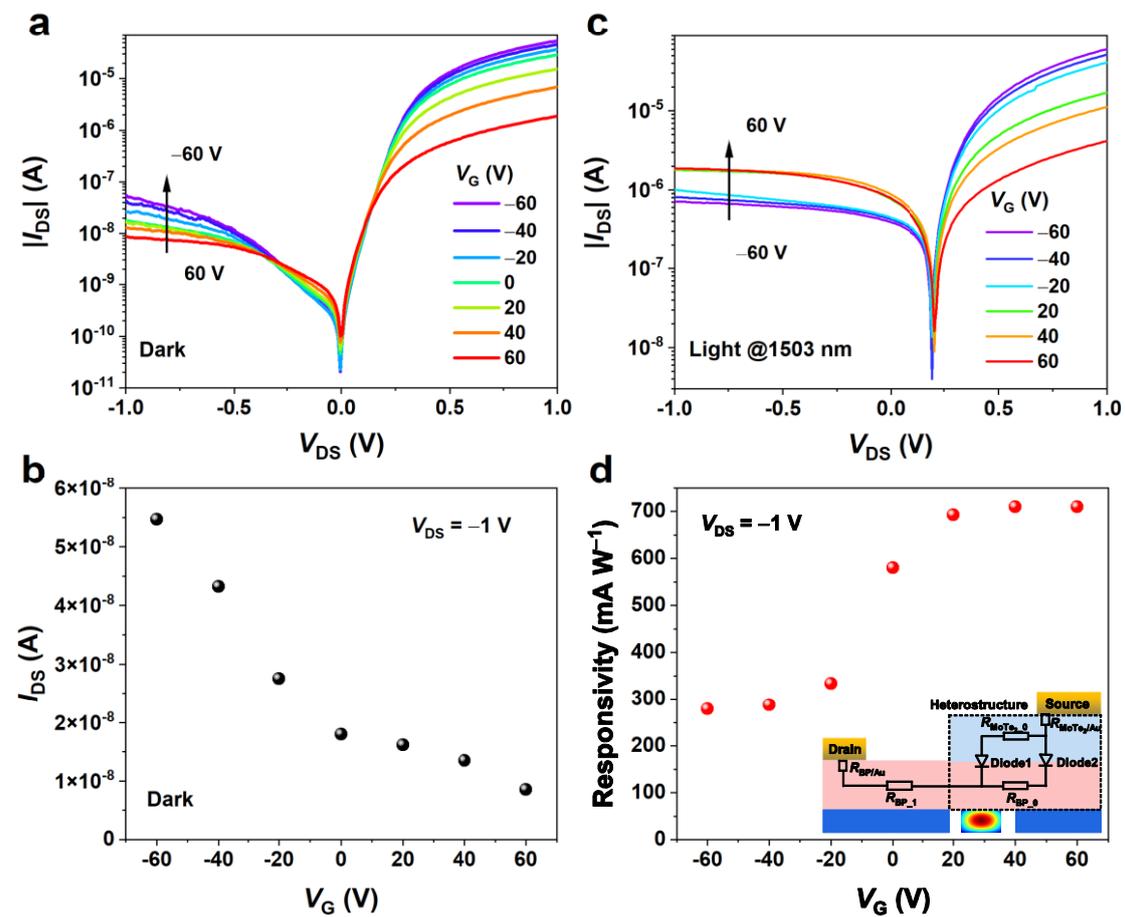

**Fig. 4 Tunable optoelectronic characteristics of the waveguide-integrated BP/MoTe$_2$ heterojunction by electrostatic gating. a** Gate tunable $I_{DS}$–$V_{DS}$ characteristics of the BP/MoTe$_2$ PN diode in darkness. **b** Dark currents of the device at $V_{DS}$ = −1 V as a function of the back gate voltage. **c** Gate tunable $I_{DS}$–$V_{DS}$ characteristics of the BP/MoTe$_2$ PN diode with the optical illumination at the wavelength of 1503 nm. **d** Intrinsic responsivities of the device at $V_{DS}$ = −1 V as a function of the back gate voltage. Inset is the cross-section schematics of the simplified equivalent circuit model of the waveguide-integrated BP/MoTe$_2$ heterojunction.

To further improve the responsivity of the waveguide-integrated BP/MoTe$_2$ heterojunction photodetetor, an out-of-plane electrostatic field is applied across it by the back gate. The BP/MoTe$_2$ vertical van der Waals heterostructure has an ultrathin thickness and a steep interfacial charge carrier gradient. The carrier density (and Fermi level) of the BP and MoTe$_2$ layers could be tuned efficiently under a vertical electrostatic field applied across them[29]. As mentioned above, the employed silicon nitride waveguide has a back substrate of doped silicon, which could be functioned as a global back-gate with an electrical voltage between it and the source electrode. Figure 4a displays the $I_{DS}$–$V_{DS}$ curves of the BP/MoTe$_2$ heterojunction measured in the darkness with different gate voltages $V_G$. When the $V_G$ increases from −60 V to 60 V, |$I_{DS}$| under positive and negative bias decreases with a drop in the rectification ratio (Fig. S8). The gate dependent dark current under $V_{DS}$ = −1 V is extracted in Fig. 4b. The dark current decreases from 54 nA to 8.5 nA. Note that, the dark current at $V_{DS}$ = −1 V and $V_G$ = 0 V obtained in this measurement is 18 nA, which is larger than the 6.8 nA obtained in Fig. 2a. It could be attributed to the degradation of the BP flake in the ambient condition during the measurements in couple days. The hBN layer is used as the upper protective layer in this work. There are some other possible methods to further improve the stability of the BP flake, such as atomic layer deposition (ALD) of dielectric layer (Al$_2$O$_3$)[40,41], PMMA upper-cladding[42], etc. Under the light illumination at a wavelength of 1503 nm, |$I_{DS}$| under positive bias shows the same trend as that in dark while |$I_{DS}$| under negative bias increase when $V_G$ increases (Fig. 4c). The increase of |$I_{DS}$| under negative bias indicates a more efficient generation of photocurrent. As shown in Fig. 4d, the intrinsic responsivity is improved from 279 mA W$^{-1}$ to 709 mA W$^{-1}$ as $V_G$ changes from −60 V to 60 V. We notice that the best intrinsic responsivity of 709 mA W$^{-1}$ is achieved at $V_G$ = 60 V and the value of the responsivity remains unchanged as $V_G$ increases, which could be limited by saturation of BP absorption[20].

The increase of responsivity and suppression of dark current when $V_G$ increases from −60 V to 60 V can be explained with a simplified equivalent circuit model (Fig. S9 or the inset of Fig. 4d). This simplified equivalent circuit model consists of several parts: the contact resistances of Au/MoTe$_2$ ($R_{MoTe2/Au}$) and Au/BP ($R_{BP/Au}$), the sheet resistance of BP ($R_{BP\_1}$, $R_{BP\_0}$) and MoTe$_2$ channels ($R_{MoTe2\_0}$), and two diodes (Diode1, Diode2). $R_{BP\_0}$ represents the sheet resistance from the BP in heterostructure area. Diode1 represents the heterostructure on the top of the waveguide while Diode2 represents the other part of the heterostructure. As shown in Fig. S10, MoTe$_2$ is n type while BP is p type. When $V_G$ increases from negative to positive values, $R_{MoTe2\_0}$ decreases while $R_{BP\_1}$ and $R_{BP\_0}$ increase. The strongly increased $R_{BP\_1}$ and $R_{BP\_0}$ could account for the decreased dark $|I_{DS}|$ measured under both conditions of positive and negative bias, as the results shown in Fig. 4a. With the optical illumination by the waveguide mode, the photogenerated electrons and holes in the BP layer will contribute to the variations of $|I_{DS}|$, as shown in Fig. 4c. Under the negative bias ($V_{DS} < 0$), $|I_{DS}|$ increases gradually as $V_G$ increases from −60 V to 60 V, which is opposite to that in darkness. In the case with a large positive $V_G$ (e.g., $V_G$ = 60 V), $R_{MoTe2\_0}$ is small and $R_{BP\_0}$ is large due to the field effect (Fig. S10). The effective voltage over the Diode1 is larger than that over the Diode2, which induces a larger built-in electric field across the Diode1 than the Diode2. Since the Diode1 is on the top of the waveguide mode, photogenerated carriers overlaps with the large built-in electric field, resulting in a large photocurrent. As $V_G$ decreases gradually to negative values, $R_{MoTe2\_0}$ ($R_{BP\_0}$) become larger (smaller). The effective voltage (built-in electric field) over the Diode1 is weakened, Hence, the photogenerated carriers in Diode1 region can not be harvested effectively, which reduces the photocurrent. Though the built-in electric field in the Diode2 is strengthened, there is no photogenerated carriers in this region. As a result, under $V_{DS} < 0$ and the optical illumination, the remarkable photocurrents strengthened by the positive $V_G$ make the $|I_{DS}|$ larger than that under the negative $V_G$. Under the positive bias ($V_{DS} > 0$), because both Diode1 and Diode2 are turned on, $|I_{DS}|$ is dominated by the dark current though there is an optical illumination. Hence, the $V_G$-dependence of $|I_{DS}|$ in Fig. 4c is same as that in Fig. 4a under $V_{DS} > 0$.

This unique gate-tunability of the waveguide-integrated 2D van der Waals heterojunction photodetector is in stark contrast to conventional bulk photodiodes integrated on a waveguide, in which the photocurrent is mainly determined by the intrinsic built-in potential. The junctions in bulk semiconductors cannot be easily modulated once their fabrications were finished.

**Discussion**

In conclusion, we proposed and demonstrated a waveguide-integrated photodetector based on a BP/MoTe$_2$ van der Waals PN heterojunction. It presents ultralow dark currents and high responsivities due to the current-rectifying characteristic of the PN heterojunction. From a fabricated device, under an external bias of 1 V pointing from n-type MoTe$_2$ to p-type BP, the intrinsic (extrinsic) responsivity is as high as 577 (283) mA W$^{-1}$ with a low dark current of 6.8 nA, leading to a remarkable NPDR of $4.13\times10^4$ mW$^{-1}$. The built-in electric field of the PN heterojunction also enables a self-driven photodetection with an intrinsic responsivity of 277 mA W$^{-1}$ under zero-bias condition. Impulse measurement shows that this kind of photodetectors can operate with a dynamic response bandwidth of 1.0 GHz. Due to the narrow bandgap of the few-layer BP, these photodetection performances are valid over a wide spectral range covering the telecommunication wavelengths. Relying on the ultrathin thickness and steep interfacial charge carrier gradient of the BP/MoTe$_2$ van der Waals heterostructure, its photodetection could be further improved by a simple electrostatic gating. An intrinsic (extrinsic) responsivity of 709 (340) mA W$^{-1}$ was obtained with a gate voltage of 60 V under an external bias of 1 V pointing from n-type MoTe$_2$ to p-type BP. Figure of merit of our phototdetector such as dark current and responsivity are on par with the reported Si-Ge waveguide photodetectors though the bandwidth of this device still needs to be optimized, as compared in Supplementary Table 2. While the silicon nitride waveguide is employed in this work, considering the layered structure and dangling-bond-free surface of the 2D materials, the proposed chip-integrated van der Waals PN heterojunction could be applied to on-chip waveguides or resonators based on other PIC materials, such as silicon, lithium niobate, polymers, etc. In addition, the employed transfer-printing technology for constructing the van der Waals PN heterojunction in our proposed device guarantees the cleanliness of the interface at the heterostructure, which does not take care to other factors, such as the twist angles of the stacked layers. Thus, it is possible to fabricate the waveguide-integrated CVD-grown van der Waals heterojunction photodetectors by using standard fabrication technology, based on the mature growth of large-scale crystalline BP and MoTe$_2$ films[43-45]. This could facilitate the development of the proposed device in future practical applications. Our work could open an avenue to develop high-performance on-chip photodetectors for various PICs with ease of fabrication, low dark current, and high responsivity.

**Materials and methods**

## Device fabrication

The employed silicon nitride waveguides and MZIs were fabricated in a 300 nm thick silicon nitride slab on a highly doped silicon substrate with a 2 μm buried oxide layer. Electron-beam lithography (EBL, NanoBeam Limited nB5) and inductively coupled plasma etching (Oxford Instruments, Plasma Pro 100 Cobra300) were used to fabricate the silicon nitride waveguides and MZIs. The interface is important for the high quality PN heterojunction. To ensure that, mechanically exfoliated h-BN, BP, MoTe$_2$ (purchased from HQ graphene) were sequentially transferred onto the silicon nitride waveguide with a dry transfer technique[30]. Au electrodes with a thickness of 50 nm were prepared on a clean silicon substrate using a shadow mask by thermal evaporation, which were then mechanically peeled and released onto the BP and MoTe$_2$ flakes[30]. The finished BP/MoTe$_2$ heterojunction was finally annealed in a tube furnace (BTF-1200C, BEQ) filled with forming gas (95% Ar and 5% H$_2$) at 200 °C for 2 h.

## Electrical and photoresponse characterizations

The electrical characteristics of the fabricated BP/MoTe$_2$ heterojunction devices were measured at room temperature under ambient conditions using a Keysight B2912A dual-channel digital source meter. A tunable laser in the telecom-band (TUNICS T100S-HP) was used as the light source. High-frequency photoresponses were performed by the impulse response measurements. A train of optical pulses (with a pulse width of 5 ps and a repetition rate of 100 MHz) at the wavelength of 1550 nm generated by an optical parametric oscillator was coupled into the device via the grating couplers. The impulse photocurrents generated from the fabricated photodetector were extracted by using a RF GS pin (MPI T26A GS150) and a bias Tee (Anritsu, G4N37, 8 kHz-40 GHz). An electrical amplifier (Mini-circuits, ZKL-1R5+, 10-1500 MHz) was used to amplify the extracted impulse photocurrents. Finally, the amplified electrical signal was transmitted into the oscilloscope (Lecroy, 740Zi-A) and the impulse response was measured.

## Acknowledgements

This project was primarily supported by the National Key R&D Program of China (Grant Nos.

2018YFA0307200 and 2017YFA0303800), the National Natural Science Foundation of China (Grant Nos. 61905198, 61775183, 11634010, and 61675171), Key Research and Development Program in Shaanxi Province of China (Grant Nos. 2017KJXX-12, 2018JM1058, and 2018KW-009), the Fundamental Research Funds for the Central Universities (Grant Nos. 3102017jc01001, 3102018jcc034, and 3102017HQZZ022).

**Author details**

[1]Key Laboratory of Light Field Manipulation and Information Acquisition, Ministry of Industry and Information Technology, and Shaanxi Key Laboratory of Optical Information Technology, School of Physical Science and Technology, Northwestern Polytechnical University, 710129 Xi'an, China. [2]Photonics Research Group, Center for Nano and Biophotonics, Ghent University, B-9000 Gent, Belgium. [3]Materials Science Factory, Instituto de Ciencia de Materiales de Madrid (ICMM-CSIC), E-28049 Madrid, Spain. [4]Department of Micro- and Nanosciences, Aalto University, Tietotie 3, FI-00076 Espoo, Finland

**Author contribution**

X. G. conceived and supervised the project. R. T. carried out the device fabrication and characterization. X. C. helped on the data analysis. C. L., S. H., and L. G. assisted the device fabrication. D. T., A. C., Z. S., and J. Z. provided valuable discussions and improvements of manuscript.

**Conflict of interest**

The authors declare that they have no conflict of interest.

**Supplementary information** is available for this paper at https://doi.org/xx.